\documentclass[useAMS,usenatbib]{mn2e}
\usepackage{amsmath}
\usepackage{rotating}

\newcommand{\lsim}{\raisebox{-3.8pt}{$\;\stackrel{\textstyle <}{\sim}\;$}}

\title[Cusp Summations and Relations of Quad Lenses]{Cusp Summations and Cusp Relations of Simple \\ Quad Lenses}
\author[Zhe Chu, G. L. Li and W. P. Lin]{Zhe Chu$^{1}$\thanks{E-mail:
chuzhe@pmo.ac.cn}, G. L. Li$^{1}$ and W. P. Lin$^{2,3}$\\
$^{1}$Purple Mountain Observatory, Chinese Academy of Sciences, 2 West Beijing Road, Nanjing 210008, China \\
$^{2}$School of Astronomy and Space Science, Sun Yat-Sen University, Guangzhou 510275, China \\
$^{3}$Shanghai Astronomical Observatory, Chinese Academy of Sciences, 80 Nandan Road, Shanghai 200030, China}
\begin{document}

\date{Accepted 2015 January 25. Received 2014 November 4; in original form 2014 October 11}

\pagerange{\pageref{firstpage}--\pageref{lastpage}} \pubyear{2014}

\maketitle

\label{firstpage}

\begin{abstract}
We review five often used quad lens models, each of which has analytical solutions and can produce four images at most. Each lens model has two parameters, including one that describes the intensity of non-dimensional mass density, and the other one that describes the deviation from the circular lens. In our recent work, we have found that the cusp and the fold summations are not equal to 0, when a point source infinitely approaches a cusp or a fold from inner side of the caustic. Based on the magnification invariant theory, which states that the sum of signed magnifications of the total images of a given source is a constant, we calculate the cusp summations for the five lens models. We find that the cusp summations are always larger than 0 for source on the major cusps, while can be larger or smaller than 0 for source on the minor cusps. We also find that if these lenses tend to the circular lens, the major and minor cusp summations will have infinite values, and with positive and negative signs respectively. The cusp summations do not change significantly if the sources are slightly deviated from the cusps. In addition, through the magnification invariants, we also derive the analytical signed cusp relations on the axes for three lens models. We find that both on the major and the minor axes the larger the lenses deviated from the circular lens, the larger the signed cusp relations. The major cusp relations are usually larger than the absolute minor cusp relations, but for some lens models with very large deviation from circular lens, the minor cusp relations can be larger than the major cusp relations.
\end{abstract}


\section{Introduction}

Producing multiple images of distant quasars or galaxies by foreground galaxies or galaxy clusters is one of the most distinct qualities of strong gravitational lensing. For nonsingular lenses, it is well known that the total image number is odd \citep{bur81}. If a point source lies within the central astroid caustic of the elliptic lens, there will be five images produced. There are two positive minima (minima point of Fermat potential, similar hereafter) images outside of the tangential critical curve, two negative saddle images inside of the tangential critical curve, and one positive maxima image lying near the lens centre \citep{bla86,sah03}. However, the maxima image located near the lens centre is usually highly demagnified and faint, resulting in four observed images.

There are some important magnification relations for the multiple image lenses. The magnification invariant means that, for some specific lens models, the sum of signed magnifications for all lensed images of a given point source is a constant, i.e., $I=\sum_{i}\mu_{i}$ \citep{dal98}. It is very interesting and surprising that the invariants are independent of most of the model parameters. For example, the magnification invariants of the point lens and Singular Isothermal Sphere (SIS) lens are 1 and 2 respectively, no matter how large the Einstein radii are and where the positions of the point sources are, as long as there are two images produced.

The cusp and fold relations are local magnification relations compared with the magnification invariant. If a point source moves to the cusp from the inner side of the tangential caustic, three of the images will merge together near the critical curve. The three close images have an asymptotic magnification relation \citep{bla86,sch92a,sch92b,mao92}
\begin{equation}
R_{\textrm{cusp}}=\frac{S_{\textrm{cusp}}}{S_{|\textrm{cusp}|}}= \frac{\mu_{\textrm{A}}+\mu_{\textrm{B}}+\mu_{\textrm{C}}}{|\mu_{\textrm{A}}|+|\mu_{\textrm{B}}|+|\mu_{\textrm{C}}|} ,
\end{equation}
where $\mu$ are the signed magnifications of the triple images A, B and C. Here, we define $S_{\textrm{cusp}}$ and $S_{|\textrm{cusp}|}$, and name the numerator $S_{\textrm{cusp}}$ \emph{cusp summation}, which will be frequently used in this work. If the point source infinitely approaches the cusp, the cusp relation $R_{\textrm{cusp}}$ will be close to 0.

A similar magnification relation holds when the source lies near a fold caustic. In this case, two images lie closely together, straddling the critical curve. One of two images is a minima and the other one is a saddle. The fold image pair also has an asymptotic magnification relation \citep{bla86,sch92b,mao92,kee05,gol10}
\begin{equation}
R_{\textrm{fold}}=\frac{S_{\textrm{fold}}}{S_{|\textrm{fold}|}}= \frac{\mu_{\textrm{A}}+\mu_{\textrm{B}}}{|\mu_{\textrm{A}}|+|\mu_{\textrm{B}}|} ,
\end{equation}
where $\mu$ are the signed magnifications of the double images A and B. Here, we define $S_{\textrm{fold}}$ \emph{fold summation} and another quantity $S_{|\textrm{fold}|}$ as before. If the source infinitely approaches the fold line, the fold relation $R_{\textrm{fold}}$ will also be close to 0.

In some previous works, when the point source infinitely approaches the cusp or the fold, the numerators in Equations (1) and (2) are also considered to be equal to 0 \citep{zak95,aaz09,pet10}. In our recent work, we (\citealt{chu13b}) proved that $S_{\textrm{cusp}}$ and $S_{\textrm{fold}}$ are usually not equal to 0. Consequently, there are different signs in the numerators, so in the definitions about the two relations we do not use the absolute value of the summed magnifications in the numerators as some other authors do.

In strong gravitational lensing, the positions of most multiple images can be fitted adequately using simple smooth lens models. Nevertheless, the observed flux ratios are more difficult to match \citep{koc91}. Actually, most of the observed fluxes of image pairs and triples disagree with the fold and cusp relations. The discrepancy between the predicted and observed flux ratios is commonly referred to as the anomalous flux ratio problem \citep{mao04,con05,mck07,shi08}. Currently the most favoured explanation of the flux ratio anomalies invokes the perturbation effects from
small-scale structures hosted by lensing galaxies \citep{mao98,met02,koc04,aaz06,mac06,che11,xu09,xu15}. In this work, based on the magnification invariant, we mainly study the cusp summation and cusp relation through five frequently used smooth quad lenses, and they may be helpful for our understanding of the anomalous flux ratio problem in another aspect.

\section{General Review for the Five Quad Lenses}

\begin{figure}
\centering
\includegraphics[width=0.4\textwidth]{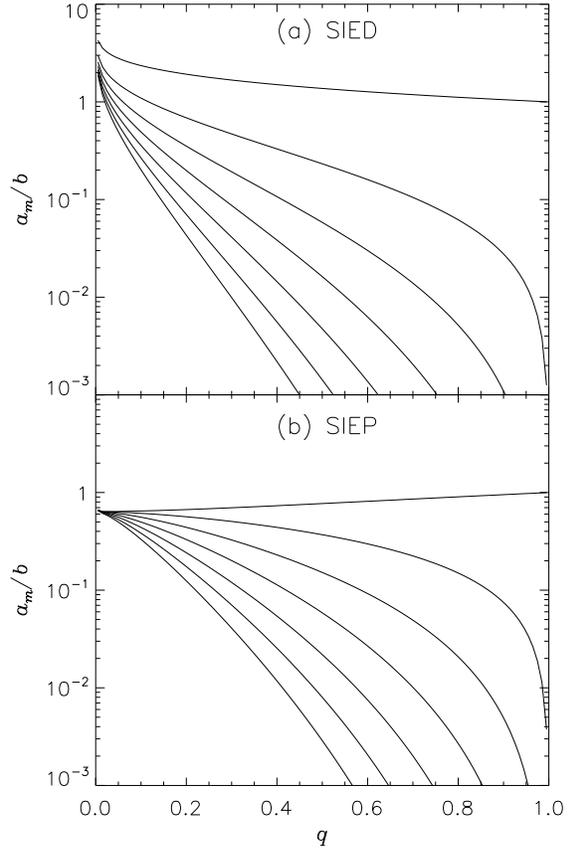}
\caption{The coefficients of multipole expansions of the SIED and SIEP lenses. For each panel, the seven curves describe the coefficients of modes $m=0$, 2, 4, 6, 8, 10, 12 from top to bottom, respectively.}
\end{figure}

We review five often used quad lenses in strong lensing, including Singular Isothermal Elliptical Density (SIED), Singular Isothermal Elliptical Potential (SIEP), Singular Isothermal Quadrupole (SIQ), SIS+shear, and Point+shear lenses. There are some similar properties for the five lenses. For each lens model, the radial critical curve degenerates into a point in the lens centre, and corresponds to the pseudo-caustic \citep{eva98}. Each of the five lenses has an astroid caustic which has four cusps and four folds, and each lens can produce four images at most for a single source.

The detailed information about these lenses is shown in Table 1. Each lens model has two parameters, including one that describes the intensity of non-dimensional mass density, and the other one that describes the deviation from the circular lens. For the two SIE lenses, they are usually studied in the Cartesian coordinates, while for the last three lens models, it is more convenient to treat them in polar coordinates.

The two SIE lenses are extended from the SIS lens. Here $b$ is used as a constant parameter, which indicates the intensity of the mass density. $q$ is the axial ratio of the SIE lens. The SIE lenses can be derived by changing $\theta$ into $\sqrt{q^{2}x^2+y^2}$ through the SIS lens. For the SIED lens the $\theta$ was changed in the mass distribution $\kappa$, while for the SIEP lens it was changed in the potential $\psi$.

\begin{table*}
 \centering
  \caption{Five simple quad lenses.}
  \begin{tabular}{@{}lcc@{}}
  \hline
  Lens model & SIED$^{a}$ $(x,y)$ & SIEP$^{b}$ $(x,y)$  \\

 \hline
 Parameters & $b,\ q\ (0<q<1)$ & $b,\ q\ (0<q\leqslant1)$  \\
 Convergence & $\kappa=\frac{b}{2\sqrt{q^{2}x^{2}+y^{2}}}$ & $\kappa=\frac{bq^2(x^{2}+y^{2})}{2(q^{2}x^{2}+y^{2})^{3/2}}$  \\
 Deflection potential & $\psi=x\alpha_{x}+y\alpha_{y}$ & $\psi=b\sqrt{q^2x^2+y^2}$  \\
 Deflection angle & $\alpha_{x}=\frac{b}{\sqrt{1-q^{2}}}\tan^{-1}\left(\sqrt{\frac{1-q^{2}}{q^{2}x^{2}+y^{2}}}x\right)$ & $\alpha_{x}=\frac{bq^2x}{\sqrt{q^{2}x^{2}+y^{2}}}$  \\
  Deflection angle & $\alpha_{y}=\frac{b}{\sqrt{1-q^{2}}}\tanh^{-1}\left(\sqrt{\frac{1-q^{2}}{q^{2}x^{2}+y^{2}}}y\right)$ & $\alpha_{y}=\frac{by}{\sqrt{q^{2}x^{2}+y^{2}}}$  \\
 Shear & $\gamma=\kappa$ & $\gamma=\kappa$  \\
 Magnification & $\mu^{-1}=1-2\kappa$ & $\mu^{-1}=1-2\kappa$   \\
 Critical curve & $q^{2}x^{2}+y^{2}=b^2$ & $\frac{bq^2(x^{2}+y^{2})}{(q^{2}x^{2}+y^{2})^{3/2}}=1$  \\
 Major axis $\beta_{\textrm{cusp}}$ & $\frac{b}{q}-\frac{b}{\sqrt{1-q^{2}}}\tan^{-1}\left(\frac{\sqrt{1-q^{2}}}{q}\right)$ & $\frac{b}{q}-bq$   \\
 Minor axis $\beta_{\textrm{cusp}}$ & $-b+\frac{b}{\sqrt{1-q^{2}}}\tanh^{-1}(\sqrt{1-q^{2}})$ & $ b-bq^2$   \\
 Naked cusp & $q \lsim 0.39$ & $q<\sqrt{2}/2$  \\
 Magnification invariant$^{f}$ & $\approx2.8$ & 2  \\

\hline
\end{tabular}
\end{table*}

\begin{table*}
 \centering
  \begin{tabular}{@{}ccc@{}}
  \hline
  SIQ$^{c}$ $(\theta,\phi)$ & SIS+shear$^{d}$ $(\theta,\phi)$ & Point+shear$^{e}$ $(\theta,\phi)$ \\

 \hline
 $\theta_{\textrm{E}},\ k\ (0\leqslant k\leqslant1)$ & $\theta_{\textrm{E}},\ \gamma\ (0\leqslant \gamma<1)$ & $\theta_{\textrm{E}},\ \gamma\ (0\leqslant \gamma<1)$ \\
 $\kappa=\frac{\theta_{\textrm{E}}}{2\theta}(1+k\cos 2\phi)$ & $\kappa=\frac{\theta_{\textrm{E}}}{2\theta}$ & $\kappa=\pi\theta_{\textrm{E}}^{2}\delta(\boldsymbol{\theta})$ \\
 $\psi=\theta_{\textrm{E}}\theta-\frac{1}{3}\theta_{\textrm{E}}k\theta\cos 2\phi$ & $\psi=\theta_{\textrm{E}}\theta-\frac{\gamma}{2}\theta^2\cos 2\phi$ & $\psi=\theta_{\textrm{E}}^2\ln\theta-\frac{\gamma}{2}\theta^2\cos 2\phi$ \\
 $\alpha_{\textrm{rad}}=\theta_{\textrm{E}}-\frac{1}{3}\theta_{\textrm{E}}k\cos 2\phi$ & $\alpha_{\textrm{rad}}=\theta_{\textrm{E}}-\gamma\theta\cos 2\phi$ & $\alpha_{\textrm{rad}}=\frac{\theta_{\textrm{E}}^2}{\theta}-\gamma\theta\cos 2\phi$ \\
 $\alpha_{\textrm{tan}}=\frac{2}{3}\theta_{\textrm{E}}k \sin 2\phi$ & $\alpha_{\textrm{tan}}=\gamma\theta\sin 2\phi$ & $\alpha_{\textrm{tan}}=\gamma\theta\sin 2\phi$ \\
 $\gamma=\kappa$ & $\boldsymbol{\gamma}=\frac{\theta_{\textrm{E}}}{2\theta}+\gamma\cos 2\phi-\textrm{i}\gamma\sin 2\phi$  & $\boldsymbol{\gamma}=\frac{\theta_{\textrm{E}}^{2}}{\theta^{2}}+\gamma\cos 2\phi-\textrm{i}\gamma\sin 2\phi$ \\
 $\mu^{-1}=1-2\kappa$ & $\mu^{-1}=1-\gamma^{2}-\frac{\theta_{\textrm{E}}}{\theta}(1+\gamma\cos 2\phi)$ & $\mu^{-1}=1-\gamma^{2}-\frac{\theta_{\textrm{E}}^{4}}{\theta^{4}}-2\frac{\theta_{\textrm{E}}^{2}}{\theta^{2}}\gamma\cos 2\phi$  \\
 $\theta=\theta_{\textrm{E}}+\theta_{\textrm{E}}k\cos 2\phi$ & $\theta=\theta_{\textrm{E}}\frac{1+\gamma\cos 2\phi}{1-\gamma^{2}}$ & $(\frac{\theta_{\textrm{E}}}{\theta})^2=\sqrt{1-\gamma^{2}\sin^{2}2\phi}-\gamma\cos2\phi$  \\
 $\beta_{\textrm{cusp}}=\frac{4}{3}\theta_{\textrm{E}}k$ & $\beta_{\textrm{cusp}}=2\theta_{\textrm{E}}\frac{\gamma}{1\mp\gamma}$ & $\beta_{\textrm{cusp}}=2\theta_{\textrm{E}}\frac{\gamma}{\sqrt{1\mp\gamma}}$  \\
 $k>0.6$ & $\gamma>1/3$ & Never  \\
 1 & $2/(1-\gamma^{2})$ & $1/(1-\gamma^{2})$ \\

\hline
\end{tabular}

 \medskip
 $^{a}$\citet{kas93,kor94,kee98,kee00}. $^{b}$\citet{kas93,wit00a}. $^{c}$\citet{koc91,wol12,chu13b}. $^{d}$\citet{kov87,fin02,kee03}. $^{e}$\citet{cha79,cha84,sch92b,an06}. $^{f}$\citet{dal98,dal01,wit00}
\end{table*}

For the SIED and SIEP lenses, their convergence can also be written in the form of polar coordinates
\begin{equation}
\kappa_{\textrm{SIED}}=\frac{1}{2\theta}\frac{b}{\sqrt{q^{2}\cos^{2}\phi+\sin^{2}\phi}}
=\frac{1}{2\theta}\frac{b\sqrt{1+\epsilon}}{\sqrt{1-\epsilon\cos 2\phi}} ,
\end{equation}
\begin{equation}
\kappa_{\textrm{SIEP}}=\frac{1}{2\theta}\frac{bq^2}{(q^{2}\cos^{2}\phi+\sin^{2}\phi)^{3/2}} =\frac{1}{2\theta}\frac{b\sqrt{1+\epsilon}(1-\epsilon)}{(1-\epsilon\cos 2\phi)^{3/2}} .
\end{equation}
Here, the parameter $\epsilon$ is related to the axial ratio $q$ by $\epsilon=(1-q^2)/(1+q^2)$.

For each SIE lens, the convergence has the form $\kappa=G_{\textrm{SIE}}(\phi)/2\theta$. The shape function $G_{\textrm{SIE}}(\phi)$ can be decomposed into multipoles through Fourier transform method \citep{kee03}
\begin{equation}
\begin{split}
G_{\textrm{SIE}}(\phi)=\sum_{m=0}^{\infty}a_m\cos m\phi , \\
a_m=\frac{1}{2\pi}\int_0^{2\pi}G_{\textrm{SIE}}(\phi)\cos m\phi\textrm{d}\phi .
\end{split}
\end{equation}
Based on the symmetry of the SIE lenses, the coefficients $a_{m}$ of the odd modes are equal to 0. Therefore, there are only even modes in the expansions, and their phases $\phi_{m}$ are all equal to 0.

Figure 1 shows the Fourier expansion coefficients of the shape functions of the SIE lenses. The strength of the monopole is the largest one among all modes. Since the Einstein radius is determined by the monopole, the Einstein radius of the SIE lens is $\theta_{\textrm{E}}=a_{0}$. For a given parameter $b$, with increasing the ratio $q$, $\theta_{\textrm{E}}$ decreases for the SIED lens, while increases for the SIEP lens. Except the monopole $a_{0}$, with increasing $q$, all the coefficients of the two SIE lenses decrease. In addition, the coefficients of all the even modes decrease with increasing of the mode $m$. The SIQ lens, also called SIS+elliptical lens \citep{dal98,wol12}, can be thought as the lowest order multipole expansion of the two SIE lenses, and its coefficients only include the monopole $a_{0}$ and the quadrupole $a_{2}$.

In polar coordinates, the shear can be decomposed into two parts, tangential or radial shear $\gamma_{+}$, and skew shear $\gamma_{\times}$ \citep{ber09}. We can write them into the complex form $\boldsymbol{\gamma}=\gamma_{+}+\textrm{i}\gamma_{\times}$. The direction of the skew shear is rotated $45^{\circ}$ from those of the tangential or radial shear. For the real part, when $\gamma_{+}>0$, it is tangential shear, while oppositely it is radial shear. The radial shear is only obvious for the void or the lens with negative mass distributions. For example, the convergence $\kappa=\cos2\phi/2\theta$, can produce tangential shear in the positive density region, and radial shear in the negative region. In addition, the external shear $\gamma$ in the last two lens models can bring both $\gamma_{+}=\gamma\cos 2\phi$ and $\gamma_{\times}=-\gamma\sin 2\phi$ components.

For any lens with a convergence of the form of $G(\phi)/2\theta$, one can decompose it into multipoles, similar to Equation (5). For each mode of this lens, except $m=1$, it has $\gamma=\kappa$ \citep{chu13a}. These shears are all tangential or radial ones $\gamma_{+}$, not including skew shear $\gamma_{\times}$. According to the superposition principle, we can conclude that, as long as the function $G(\phi)$ do not include the $m=1$ mode, the shear of the lens is $\gamma=\kappa$, and the magnification is $\mu=1/(1-2\kappa)$. In fact, \citet{wit00a} found that the lens potential with $\psi=\theta F(\phi)$ form has the magnification of $\mu=1/(1-2\kappa)$. It is consistent with our conclusion, because for a convergence $G(\phi)/2\theta$ with no monopole, the lens potential can be written in the form $\psi=\theta F(\phi)$ \citep{eva01,eva03}.

The last two lenses are derived by adding a uniform external shear on the SIS or the point lens. The point+shear lens is also called Chang-Refsdal lens \citep{cha79,cha84}. There is usually a minus sign before $\gamma$ in few other studies. We use it in this form to let the major axis lie along the \emph{X}-axis, and minor axis along the \emph{Y}-axis. For a general strong lens, when the $\theta$ is infinite large, the magnification $\mu$ should have a positive sign. Therefore, it needs the external shear $\gamma$ to be smaller than 1. However, it could also be studied with $\gamma>1$ for extreme cases, as in \citet{an06}.

In addition, the convergence $\pi\theta_{\textrm{E}}^{2}\delta(\boldsymbol{\theta})$ of the point lens or the point+shear lens is derived through the relation of the two-dimensional Dirac delta function $\nabla^{2}\ln|\boldsymbol{\theta}|=2\pi\delta(\boldsymbol{\theta})$. Here for the point lens, we do not consider the Schwarzschild radius of the point mass, near which the deflection angles are more complex \citep{vir00}.

From the deflection angles and critical curves one can easily calculate the angular distance $\beta_{\textrm{cusp}}$ from the cusp to the source centre, and the angular distance $\beta_{\textrm{pseu}}$ from pseudo-caustic to the source centre on the axes. When the two quantities are equal to each other, we can derive the critical value of the second parameter for naked cusp appearing.

The lens equation $\boldsymbol{\beta}=\boldsymbol{\theta}-\boldsymbol{\alpha}$ includes two independent equations. It describes the transformation between the lens plane $(\theta, \phi)$ and the source plane $(\beta, \varphi)$, and can also be written in polar coordinates as
\begin{equation}
\beta^2=(\theta-\alpha_{\textrm{rad}})^2+\alpha_{\textrm{tan}}^2 ,
\end{equation}
\begin{equation}
\tan(\phi-\varphi)=\frac{\alpha_{\textrm{tan}}}{\theta-\alpha_{\textrm{rad}}}
\end{equation}
\citep{chu13a}. Using these lens equations in the Cartesian coordinates or the polar coordinates, together with the critical curves, one can also calculate the caustics for the SIE lenses or the other three lenses.

\section{The Cusp Summations for Source on the Cusps}

\begin{figure*}
\centering
\includegraphics[width=0.9\textwidth]{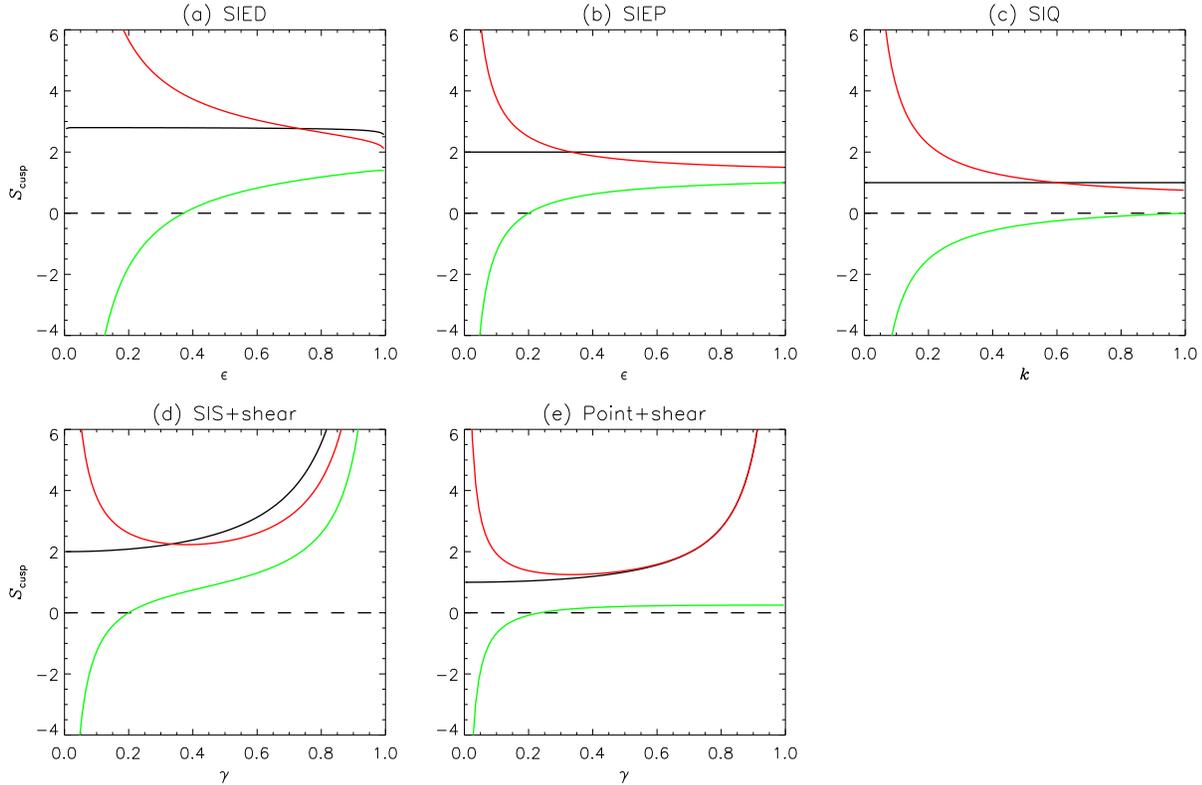}
\caption{The red and green curves show the cusp summations on the major and minor cusps, respectively. The black curves mean the magnification invariants.}
\end{figure*}

As we know, when a point source is exactly on the cusp, the cusp summation is usually not equal to 0. However, it is difficult to derive the magnifications of the triple images related to cusp summation using traditional methods. Fortunately, we have the magnification invariants for the five lens models. Therefore, we can derive the cusp summation through the differences between magnification invariant and the finite magnification of the fourth image. The magnification invariants of the SIEP, SIQ, SIS+shear, Point+shear lenses have been calculated by \citet{dal98} and \citet{dal01}. The magnification invariant of SIED lens is given by \citet{wit00}. These magnification invariants are only valid when four images are produced.

For the SIED lens, the magnification invariant is \citep{wit00}
\begin{equation}
I_{\textrm{SIED}}=\frac{2}{1-u/\tan^{-1}u}+\frac{2}{1-v/\tanh^{-1}v}\approx 2.8 ,
\end{equation}
where $u=\sqrt{2\epsilon/(1-\epsilon)}$, and $v=\sqrt{2\epsilon/(1+\epsilon)}$. The magnification invariant slightly depends on the $q$, and also slightly depends on the position of the source.

For a point source on the cusp, it has one image with finite magnification, and three images merged together with infinite magnifications. After we derive the finite magnification of the point source, together with the magnification invariant, we can get the cusp summations of the triplets both on the major and minor cusps
\begin{equation}
\begin{split}
S_{\textrm{cusp}}\approx \frac{2+u/\tan^{-1}u}{2-2u/\tan^{-1}u}+\frac{2}{1-v/\tanh^{-1}v} , \\
S_{\textrm{cusp}}\approx \frac{2}{1-u/\tan^{-1}u}+\frac{2+v/\tanh^{-1}v}{2-2v/\tanh^{-1}v} .
\end{split}
\end{equation}
When $\epsilon\approx0.73$ or $q\approx0.39$, the cusp summation of the major cusp is equal to the magnification invariant. When $\epsilon\approx0.37$ or $q\approx0.68$, the cusp summation of the minor cusp is equal to 0.

Similarly, for the SIEP lens, we can also derive the cusp summations of the triplets on the major and minor cusps
\begin{equation}
\begin{split}
S_{\textrm{cusp}}=\frac{3-2q^2}{2-2q^2}=\frac{1}{4}(5+\frac{1}{\epsilon}) , \\
S_{\textrm{cusp}}=\frac{2-3q^2}{2-2q^2}=\frac{1}{4}(5-\frac{1}{\epsilon}) .
\end{split}
\end{equation}
When $\epsilon=0.2$ or $q=\sqrt{6}/3$, the cusp summation of the minor cusp is equal to 0. For the SIQ lens, the cusp summation of the triplet on the major/minor cusp is
\begin{equation}
S_{\textrm{cusp}}=\frac{3}{8}(1\pm\frac{1}{k}) .
\end{equation}
For the SIS+shear lens, the cusp summation of the triplet on the major/minor cusp is
\begin{equation}
S_{\textrm{cusp}}=\frac{5\gamma\pm1}{4\gamma(1-\gamma^{2})} .
\end{equation}
When $\gamma=0.2$, the cusp summation of the minor cusp is equal to 0. For the Point+shear lens, the cusp summation of the triplet on the major/minor cusp is
\begin{equation}
S_{\textrm{cusp}}=\frac{4\gamma\pm(1-\gamma^{2})}{8\gamma(1\mp\gamma)} .
\end{equation}
When $\gamma\approx0.24$, the cusp summation of the minor cusp is equal to 0. When $\gamma$ approaches 1, the minor cusp summation will be close to 0.25.

For each of these lenses, the cusp summation does not depend on the first parameter of the lens, and only depends on the second parameter which describes the deviation from the circular lens. Figure 2 shows the cusp summations on the cusps for the five lens models based on Equations (9)-(13). Apparently, the cusp summations $S_{\textrm{cusp}}$ are usually not equal to 0, and are small quantities compared to the infinite magnifications of the triple images. Therefore, the value of $S_{\textrm{cusp}}$ is very easily to be ignored as higher order infinitesimals, when treat with the magnifications the three images \citep{sch92a,zak95}.

If these lenses tend to the circular lens, the major and minor cusp summations will have infinite values. The cusp summations of the major cusps are always larger than 0. When it is smaller than the magnification invariant, it means naked cusps appear. For the Point+shear lens, naked cusp will never appear, because the pseudo-caustic of this lens model is a circle with infinite large radius. In addition, unlike the major ones, the cusp summations of the minor cusps can be larger or smaller than 0 except for the SIQ lens.

\section{The Cusp Summations and Relations for Source on the Axes}

Through the difference between magnification invariant and the magnification of the fourth image, one can also calculate the cusp summation on the major or the minor axis. For each lens model, we define $z=\beta/\beta_{\textrm{cusp}}$ both on the major and minor axes. Because the analytical solutions for the SIED and point+shear lenses are very complex, we do not study them here.

For the SIEP lens, the cusp summations on the major and minor axes are
\begin{equation}
\begin{split}
S_{\textrm{cusp}}=1+\frac{1}{(1-q^{2})(1+z)} , \\
S_{\textrm{cusp}}=1-\frac{q^{2}}{(1-q^{2})(1+z)} .
\end{split}
\end{equation}
For the SIQ lens, the cusp summations on the major and minor axes are
\begin{equation}
\begin{split}
S_{\textrm{cusp}}=\frac{3(k+1)}{4k(1+z)} , \\
S_{\textrm{cusp}}=\frac{3(k-1)}{4k(1+z)} .
\end{split}
\end{equation}
For the SIS+shear lens, the cusp summations on the major and minor axes are
\begin{equation}
\begin{split}
S_{\textrm{cusp}}=\frac{\gamma(3+2z)+1}{2\gamma(1-\gamma^{2})(1+z)} , \\
S_{\textrm{cusp}}=\frac{\gamma(3+2z)-1}{2\gamma(1-\gamma^{2})(1+z)} .
\end{split}
\end{equation}

Figure 3 shows the cusp summations of the three lens models based on Equations (14)-(16). We can find that the cusp summations on the cusps are the smallest ones for source on the major axes, while the cusp summations on the cusps are the largest ones for source on the minor axes. The cusp summations do not change very much if the sources are slightly deviated from the cusps.

\begin{figure}
\centering
\includegraphics[width=0.45\textwidth]{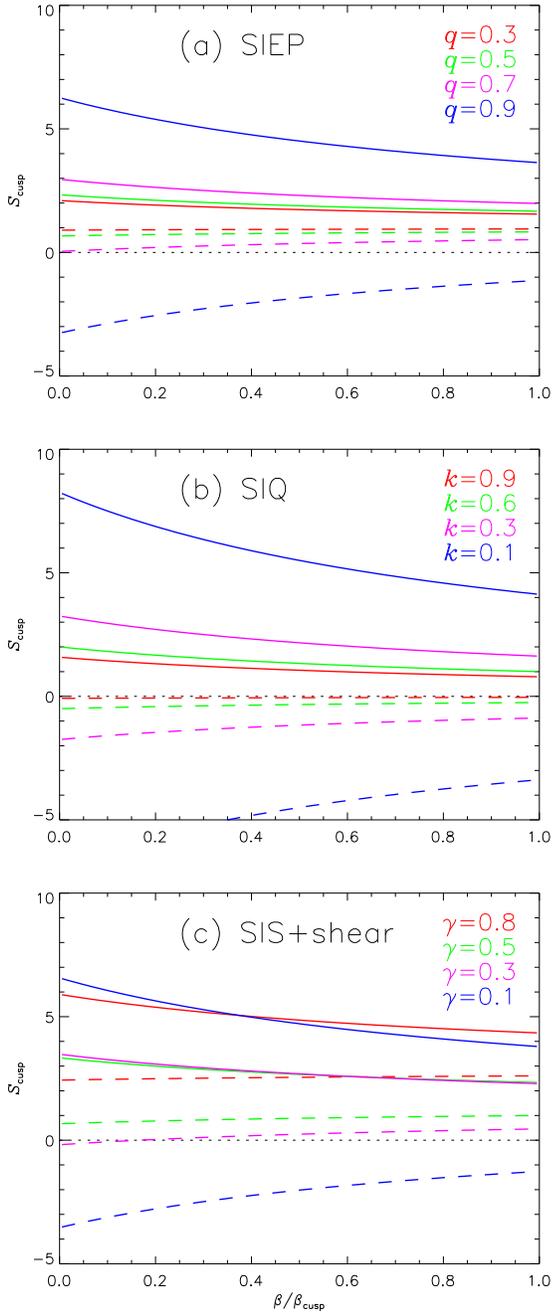}
\caption{The solid and dashed curves show the cusp summations on the major and minor axes, respectively.}
\end{figure}

If a point source lies on the axes, one can easily calculate the positions and the magnifications of the two images locating on the axes of the lens plane. However, through traditional methods, it is difficult to calculate the magnifications of the other two images which have same magnification values. Now, since we know the magnification invariants for these lens models, we can also analytically derive the magnifications of the two images lying off the axes.

After deriving the magnifications of the two side images, we can also calculate the cusp relations on the major and minor axes. For the SIEP lens, the cusp relations on the major and minor axes are
\begin{equation}
\begin{split}
R_{\textrm{cusp}}=\frac{4}{2+q^{2}+z+z^{2}-q^{2}z^{2}}-1 , \\
R_{\textrm{cusp}}=\frac{4q^{2}}{-1-2q^{2}-q^{2}z+z^{2}-q^{2}z^{2}}+1 .
\end{split}
\end{equation}
For the SIQ lens, the cusp relations on the major and minor axes are
\begin{equation}
\begin{split}
R_{\textrm{cusp}}=\frac{3(1+k)(1-z)}{9+k+3z+3kz+8kz^2} , \\
R_{\textrm{cusp}}=\frac{3(1-k)(1-z)}{-9+k-3z+3kz+8kz^2} .
\end{split}
\end{equation}
For the SIS+shear lens, the cusp relations on the major and minor axes are
\begin{equation}
\begin{split}
R_{\textrm{cusp}}=\frac{4+4\gamma}{3+\gamma+z+\gamma z +2\gamma z^2}-1 , \\
R_{\textrm{cusp}}=\frac{4-4\gamma}{-3+\gamma-z+\gamma z +2\gamma z^2}+1 .
\end{split}
\end{equation}

Figure 4 shows the cusp relations of the three lens models based on Equations (17)-(19). For each of the three lens models, the cusp relation also does not depend on the first parameter of the lens, and only depends on the second parameter. For each lens, major cusp relations approach 0 only from positive values. However, for the minor cusp relation, they can approach 0 from positive or negative values. In fact, the early work \citet{kee03} also implied that the situation for minor cusp is not certain, and regretfully they studied the absolute value of the $R_{\textrm{cusp}}$. Nevertheless, the minor cusp relations of the SIQ lens only approaches 0 from negative values.

As shown in Figures 3 and 4, on the major axes, the changes of $S_{\textrm{cusp}}$ and $R_{\textrm{cusp}}$ against the second parameters are usually opposite. E.g., in Figure 3(a), the solid red curve is the lowest one, while in Figure 4(a), the solid red curve is the highest one. On the minor axes, the changes of $S_{\textrm{cusp}}$ and $R_{\textrm{cusp}}$ are similar. It has been proved that, when $\epsilon$, $k$, and $\gamma$ are close to 0, the cusp summations $S_{\textrm{cusp}}$ are infinitely large. Through Equations (17)-(19), we find even these three parameters equal to 0, as long as $z=1$, the cusp relations $R_{\textrm{cusp}}$ are still equal to 0. It means that when these lenses tend to circular lens, $S_{|\textrm{cusp}|}$ are higher order infinities compared to the $S_{\textrm{cusp}}$.

In observations, for the major cusp relation with a positive sign, the flux summation of the two side images is larger than that of the middle image. For the minor cusp relation with a positive sign, the flux of the middle image is larger than the sum value of two side images, while with a negative sign, the sum flux of the two side images is larger than that of the middle image, which is similar to the major cusp in observation. From Figure 4, we find that the larger the lens deviated from circular lens, the larger the signed cusp relations, for both on the major and minor axes. The minor cusp relations are more sensitive to the second parameters than the major cusp relations, especially for the SIEP and SIS+shear lenses. The minor cusp relations of these two lens models can be larger than 0, and can be even larger than the major cusp relations.

The discriminant for triple images in being a major or a minor cusp type can be found in the distances of the images from the lens centre. If the distance from the triplet to lens centre is larger than that from the singlet, it is a major cusp type, otherwise it is a minor cusp type \citep{met12}. RXJ0911+0551 \citep{bad97,bru98} is the only cusp type lens system whose source lying near the minor cusp, and the signed cusp relation of this lens system is $R_{\textrm{cusp}}=-0.192\pm0.011$ (\citealt{kee03}, the negative sign is given by us). If it can be fitted by the SIEP lens, we can conclude that the axes ratio $q$ is very large. There is no sample in which the middle image has larger flux than the sum of two side images observed by now. If it can be find, it can be confirmed as minor cusp type with a larger deviation from circular lens.

\begin{figure}
\centering
\includegraphics[width=0.45\textwidth]{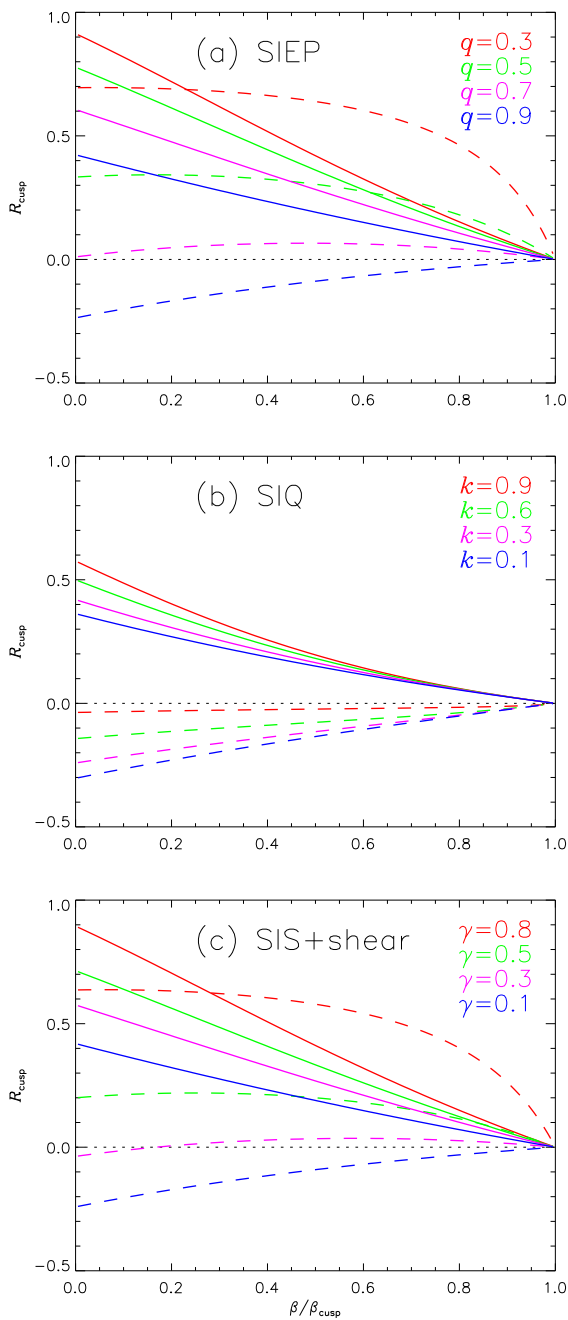}
\caption{The solid and dashed curves show the cusp relations on the major and minor axes, respectively.}
\end{figure}

\section{Conclusions and Discussion}

The four-image lens systems are very important and are very common in the observations of lensed quasars \citep{rus01,cla02}. We review five quad lens models, each of which has analytical solutions and can produce four images at most. Each of the five lenses has two parameters, including one that describes the intensity of the mass distribution, and the other one that describes the deviation from the circular lens. Using the magnification invariants of these lens models, we calculate the cusp summations for the five lenses. We find that for a point source on the cusp, the cusp summation is always larger than 0 for the major cusp, while can be larger or smaller than 0 for the minor cusp. If these lenses tend to the circular lens, the major and minor cusp summations will have infinite values, and with positive and negative signs respectively.

In this study, we calculate the cusp summations on the axes for SIEP, SIQ, SIS+shear lenses, and find that the cusp summations on the cusps are the smallest ones for sources on the major axes, while the cusp summations on the cusps are the largest ones for sources on the minor axes. The cusp summations do not change very much if the sources are slightly deviated from the cusps.

In addition, through the magnification invariants, we also calculate the magnifications of the two side images for source on the axes, and then derive the signed cusp relations on the axes. We find both on the major and the minor axes that the more the lenses deviated from the circular lens, the larger the signed cusp relations. When the point source moves to the minor cusp infinitely, the minor cusp relation has two ways to approach 0, i.e., from positive or from negative value. The changes of $S_{\textrm{cusp}}$ and $R_{\textrm{cusp}}$ against the second parameter are usually opposite for sources on the major axes. When these lenses tend to circular lens, $S_{|\textrm{cusp}|}$ are higher order infinities compared to the $S_{\textrm{cusp}}$.

The analytical results show that, the major cusp relations are usually larger than the minor cusp relations, but for some lens models with larger deviation from circular lens, the minor cusp relation can be larger than the major cusp relation. In some previous numerical work, the major cusp relations are much more easily larger than the minor cusp relations \citep{bra04,ama06,xu09,met12}. We guess that in most of the numerical simulations the ellipticity of projected haloes is too small to let the minor cusp relation to be much larger.

The cusp summation and relation do not depend on the first parameter of the lens. Therefore, if we change the redshifts of the lens body or the source (which is equivalent to multiply a constant on the first parameter), it usually does not change the cusp summation or the cusp relation. We guess the fold summation and fold relation are similar to those properties of the cusp. It should be noted that, all the conclusions in this work are based on these simple lens models, and we do not consider the nonsingular core or substructures in the real lens bodies. There is no doubt that they can influence the cusp or fold magnification relations significantly.

In future, based on the magnification invariants, we can also calculate the fold summations of these quad lenses for source exactly on the fold lines, through the finite magnifications of the other two images which can be derived using numerical method. We expect the summed magnification of fold image pair to change continuously along the fold line. Furthermore, closer to the major cusp, the fold summations will be smaller, while closer to the minor cusp, the summations will be larger, and the fold summations will have infinite values with negative and positive signs near the major and minor cusps respectively \citep{chu13b}.

Based on the cusp summations of the five lenses, we can bring forward a question: is the major cusp summation always larger than 0 for any type of major cusps? Or in other word, is it independent of the lens models? This is a very interesting mathematical problem. Solving this question can help us to understand the singularity theory or caustic metamorphoses in strong gravitational lensing. In addition, it is also very interesting to study the magnification summations for a point source on the higher order singularities of the caustics \citep{pet01,wer09,aaz11}. In these cases, more than three images with infinite magnifications will merge together, and the dependence of the magnification summations on the lens parameters are more complex than those of the cusp.

\section*{Acknowledgements}

The authors are grateful to Emanuele Contini for improving the manuscript. We acknowledge the supports by the National Key Basic Research Program of China (no. 2015CB857000) and the ``Strategic Priority Research Program the Emergence of Cosmological Structures" of the Chinese Academy of Sciences (no. XDB09000000). Z.C. is supported by National Natural Science Foundation of China (11403103) and China Postdoctoral Science Foundation (2014M551681). G.L. is supported by the One-Hundred-Talent fellowships of CAS and by the NSFC grants (11273061 and 11333008). W.P.L. acknowledges supports by the NSFC projects (11473053, 11121062, 11233005, U1331201).

\label{lastpage}


\begin{thebibliography}{99}


\bibitem[\protect\citeauthoryear{Aazami \& Natarajan}{2006}]{aaz06} Aazami A.B., Natarajan P., 2006, MNRAS, 372, 1692
\bibitem[\protect\citeauthoryear{Aazami \& Petters}{2009}]{aaz09} Aazami A.B., Petters A.O., 2009, JMP, 50, 032501
\bibitem[\protect\citeauthoryear{Aazami, Petters \& Rabin}{2011}]{aaz11} Aazami A.B., Petters A.O., Rabin J.M., 2011, JMP, 52, 022501
\bibitem[\protect\citeauthoryear{Amara et al.}{2006}]{ama06} Amara A., Metcalf R.B., Cox T.J., Ostriker J.P., 2006, MNRAS, 367, 1367
\bibitem[\protect\citeauthoryear{An \& Evans}{2006}]{an06} An J.H., Evans N.W., 2006, MNRAS, 369, 317
\bibitem[\protect\citeauthoryear{Bade et al.}{1997}]{bad97} Bade N., Siebert J., Lopez S., et al., 1997, A\&A, 317, L13
\bibitem[\protect\citeauthoryear{Bernstein \& Nakajima}{2009}]{ber09} Bernstein G.M., Nakajima R., 2009, ApJ, 693, 1508
\bibitem[\protect\citeauthoryear{Blandford \& Narayan}{1986}]{bla86} Blandford R., Narayan R., 1986, ApJ, 310, 568
\bibitem[\protect\citeauthoryear{Brada\v{c} et al.}{2004}]{bra04} Brada\v{c} M., Schneider P., Lombardi M., et al., 2004, A\&A, 423, 797
\bibitem[\protect\citeauthoryear{Burke}{1981}]{bur81} Burke W., 1981, ApJ, 244, L1
\bibitem[\protect\citeauthoryear{Burud et al.}{1998}]{bru98} Burud I., Courbin F., Lidman C., et al., 1998, ApJ, 501, L5
\bibitem[\protect\citeauthoryear{Chang \& Refsdal}{1979}]{cha79} Chang K., Refsdal S., 1979, Nature, 282, 561
\bibitem[\protect\citeauthoryear{Chang \& Refsdal}{1984}]{cha84} Chang K., Refsdal S., 1984, A\&A, 132, 168
\bibitem[\protect\citeauthoryear{Chen, Koushiappas \& Zentner}{2011}]{che11} Chen J., Koushiappas S.M., Zentner A.R., 2011, ApJ, 741, 117
\bibitem[\protect\citeauthoryear{Chu et al.}{2013}]{chu13a} Chu Z., Lin W.P., Li G.L., Kang X., 2013, ApJ, 765, 134
\bibitem[\protect\citeauthoryear{Chu, Lin \& Yang}{2013}]{chu13b} Chu Z., Lin W.P., Yang X., 2013, ApJ, 770, L34
\bibitem[\protect\citeauthoryear{Claeskens \& Surdej}{2002}]{cla02} Claeskens J.-F., Surdej J., 2002, A\&A Rev., 10, 263
\bibitem[\protect\citeauthoryear{Congdon \& Keeton}{2005}]{con05} Congdon A.B., Keeton C.R., 2005, MNRAS, 364, 1459
\bibitem[\protect\citeauthoryear{Dalal}{1998}]{dal98} Dalal N., 1998, ApJ, 509, L13
\bibitem[\protect\citeauthoryear{Dalal \& Rabin}{2001}]{dal01} Dalal N., Rabin J.M., 2001, JMP, 42, 1818
\bibitem[\protect\citeauthoryear{Evans \& Wilkinson}{1998}]{eva98} Evans N.W., Wilkinson M.I., 1998, MNRAS, 296, 800
\bibitem[\protect\citeauthoryear{Evans \& Witt}{2001}]{eva01} Evans N.W., Witt H.J., 2001, MNRAS, 327, 1260
\bibitem[\protect\citeauthoryear{Evans \& Witt}{2003}]{eva03} Evans N.W., Witt H.J., 2003, MNRAS, 345, 1351
\bibitem[\protect\citeauthoryear{Finch et al.}{2002}]{fin02} Finch T.K., Carlivati L.P., Winn J.N., Schechter P.L., 2002, ApJ, 577, 51
\bibitem[\protect\citeauthoryear{Goldberg et al.}{2010}]{gol10} Goldberg D.M., Chessey M.K., Harris W.B., Richards G.T., 2010, ApJ, 715, 793
\bibitem[\protect\citeauthoryear{Kassiola \& Kovner}{1993}]{kas93} Kassiola A., Kovner I., 1993, ApJ, 417, 450
\bibitem[\protect\citeauthoryear{Keeton \& Kochanek}{1998}]{kee98} Keeton C.R., Kochanek C.S., 1998, ApJ, 495, 157
\bibitem[\protect\citeauthoryear{Keeton, Mao \& Witt}{2000}]{kee00} Keeton C.R., Mao S., Witt H.J., 2000, ApJ, 537, 697
\bibitem[\protect\citeauthoryear{Keeton, Gaudi \& Petters}{2003}]{kee03} Keeton C.R., Gaudi B.S., Petters A.O., 2003, ApJ, 598, 138
\bibitem[\protect\citeauthoryear{Keeton, Gaudi \& Petters}{2005}]{kee05} Keeton C.R., Gaudi B.S., Petters A.O., 2005, ApJ, 635, 35
\bibitem[\protect\citeauthoryear{Kochanek}{1991}]{koc91} Kochanek C.S., 1991, ApJ, 373, 354
\bibitem[\protect\citeauthoryear{Kochanek \& Dalal}{2004}]{koc04} Kochanek C.S., Dalal N., 2004, ApJ, 610, 69
\bibitem[\protect\citeauthoryear{Kormann, Schneider \& Bartelmann}{1994}]{kor94} Kormann R., Schneider P., Bartelmann M., 1994, A\&A, 284, 285
\bibitem[\protect\citeauthoryear{Kovner}{1987}]{kov87} Kovner I., 1987, ApJ, 312, 22
\bibitem[\protect\citeauthoryear{Macci\`{o} \& Miranda}{2006}]{mac06} Macci\`{o} A.V., Miranda M., 2006, MNRAS, 368, 599
\bibitem[\protect\citeauthoryear{Mao}{1992}]{mao92} Mao S., 1992, ApJ, 389, 63
\bibitem[\protect\citeauthoryear{Mao \& Schneider}{1998}]{mao98} Mao S., Schneider P., 1998, MNRAS, 295, 587
\bibitem[\protect\citeauthoryear{Mao et al.}{2004}]{mao04} Mao S., Jing Y., Ostriker J.P., Weller, J., 2004, ApJ, 604, L5
\bibitem[\protect\citeauthoryear{McKean et al.}{2007}]{mck07} McKean J.P., Koopmans L.V.E., Flack C.E., et al., 2007, MNRAS, 378, 109
\bibitem[\protect\citeauthoryear{Metcalf \& Amara}{2012}]{met12} Metcalf R.B., Amara A., 2012, MNRAS, 419, 3414
\bibitem[\protect\citeauthoryear{Metcalf \& Zhao}{2002}]{met02} Metcalf R.B., Zhao H., 2002, ApJ, 567, L5
\bibitem[\protect\citeauthoryear{Petters, Levine \& Wambsganss}{2001}]{pet01} Petters A.O., Levine H., Wambsganss J., 2001, Singularity Theory and Gravitational Lensing. Birkh\"{a}user, Boston
\bibitem[\protect\citeauthoryear{Petters \& Werner}{2010}]{pet10} Petters A.O., Werner M.C., 2010, GReGr, 42, 2011
\bibitem[\protect\citeauthoryear{Rusin \& Tegmark}{2001}]{rus01} Rusin D., Tegmark M., 2001, ApJ, 553, 709
\bibitem[\protect\citeauthoryear{Saha \& Williams}{2003}]{sah03} Saha P., Williams L.L.R., 2003, AJ, 125, 2769
\bibitem[\protect\citeauthoryear{Schneider \& Weiss}{1992}]{sch92a} Schneider P., Weiss A., 1992, A\&A, 260, 1
\bibitem[\protect\citeauthoryear{Schneider, Ehlers \& Falco}{1992}]{sch92b} Schneider P., Ehlers J., Falco E.E., 1992, Gravitational Lenses. Springer, Berlin
\bibitem[\protect\citeauthoryear{Shin \& Evans}{2008}]{shi08} Shin E.M., Evans N.W., 2008, MNRAS, 385, 2107
\bibitem[\protect\citeauthoryear{Virbhadra \& Ellis}{2000}]{vir00} Virbhadra K.S., Ellis G.F.R., 2000, Phys. Rev. D, 62, 084003
\bibitem[\protect\citeauthoryear{Werner}{2009}]{wer09} Werner M.C. 2009, JMP, 50, 082504
\bibitem[\protect\citeauthoryear{Witt, Mao \& Keeton}{2000}]{wit00a} Witt H.J., Mao S., Keeton C.R., 2000, ApJ, 544, 98
\bibitem[\protect\citeauthoryear{Witt \& Mao}{2000}]{wit00} Witt H.J., Mao S., 2000, MNRAS, 311, 689
\bibitem[\protect\citeauthoryear{Woldesenbet \& Williams}{2012}]{wol12} Woldesenbet A.G., Williams L.L.R., 2012, MNRAS, 420, 2944
\bibitem[\protect\citeauthoryear{Xu et al.}{2009}]{xu09} Xu D.D., Mao S., Wang J., et al., 2009, MNRAS, 398, 1235
\bibitem[\protect\citeauthoryear{Xu et al.}{2015}]{xu15} Xu D., Sluse D., Gao L., Wang J., et al., 2015, MNRAS, 447, 3189
\bibitem[\protect\citeauthoryear{Zakharov}{1995}]{zak95} Zakharov A.F., 1995, A\&A, 293, 1

\end{thebibliography}
\end{document}